\documentclass[12pt,a4paper]{article}
\usepackage{latexsym}
\usepackage{epsf}
\usepackage{amssymb}
\makeatletter
\@addtoreset{equation}{section}
\makeatother

\begin{document}

\begin{flushright}
\small
IFT-UAM/CSIC-03-43\\
October, $2003$
\normalsize
\end{flushright}

\begin{center}


\vspace{.7cm}

{\Large {\bf   On  Geometric Engineering of   $N=1$  $ADE$  Quiver  
 Models }}\footnote{ 
{\tt  
 Talk presented at the Workshop on Quantum Field Theory,  Geometry  and Non 
Perturbative Physics, Rabat, July 28th-29th, 2003.}}

\vspace{.7cm}


{\bf\large Adil Belhaj}\footnote{ 
{\tt adil.belhaj@uam.es}}

\vskip 0.4truecm

\vskip 0.2cm

{\it Instituto de F\'{\i}sica Te\'orica, C-XVI,
Universidad Aut\'onoma de Madrid \\
E-28049-Madrid, Spain}

\vskip 0.2cm

\vspace{.7cm}


{\bf Abstract}

\end{center}

\begin{quotation}

\small

In this talk, we discuss  four-dimensional $N=1$  affine $ADE$  quiver gauge
models  using the 
geometric engineering  method in  M-theory on $G_2$ manifolds with 
K3 
fibrations. 

\end{quotation}

\newpage

\pagestyle{plain}

  \def\be{\begin{equation}}
  \def\ee{\end{equation}}
  \def\bea{\begin{eqnarray}}
  \def\eea{\end{eqnarray}}
  \def\nn{\nonumber}
  \def\l{\lambda}
  \def\t{\times}
  \def\[{\bigl[}
  \def\]{\bigr]}
  \def\({\bigl(}
  \def\){\bigr)}
  \def\p{\partial}
  \def\o{\over}
  \def\ta{\tau}
  \def\cm{\cal M}
  \def\R{\bf R}
  \def\b{\beta}
  \def\a{\alpha}

\section{Introduction}

A well-known way to get supersymmetric quantum field theory (QFT)  
from superstrings, M- or F-theory, is to consider compactifications on  
a singular manifold $X$ with K3 fibration over a base space 
$B$. This method is called geometric engineering 
\cite{KV, KLMVW,KaV,KKV,BJPSV, KMV,BFS, ABS}. In this way, the gauge group 
$G$ and matter content 
of the QFT are defined by the singularities of the fiber and 
the non-trivial geometry of the base space, respectively. 
The gauge coupling $g$ is proportional to the inverse of 
the square root of the volume of the base: $V(B)=g^{-2}$.
The complete set of physical parameters of the QFT is related to 
the  geometric moduli space of the internal manifold. 
 For instance, several exact results for the Coulomb branch of  
$N=2$  four-dimensional QFT embedded in type IIA superstring theory are 
obtained naturally
by  the toric geometry realization and local mirror symmetry of Calabi-Yau 
threefolds.  
The latter are realized as a K3 fibration with $ADE$ singularity over
a projective $\bf P^1$ complex curve 
or a collection of intersecting $\bf P^1$ curves. The corresponding  $N=2$ 
QFT$_4$ 
are represented by 
quiver diagrams similar to Dynkin diagrams.  One should distinguish three 
cases:  ordinary, affine and indefinite Lie algebras \cite{ KMV,BFS, ABS}.

Quite recently, four-dimensional  gauge theories preserving only 
four supercharges have attracted a lot of attention. Work has been done 
using  either  intersecting  D-brane physics 
\cite{CSU,HI,FHHI,FH,U,CIM}, geometric transition in type II superstrings 
on  the  conifold \cite{ V, AMV, CIV, CKV, CFIKV, AGMV}, or M-theory on $G_2$ 
manifolds 
\cite{H,Be,BDR}.  \par  The aim of this talk is to discuss  $ N=1$ $ADE$  
quiver 
gauge 
models   from  M-theory compactification on  a seven-dimensional 
manifold with  $G_2$ holonomy group. 
 The manifold  is realized explicitly as K3 fibrations over a 
three-dimensional base
space with Dynkin geometries.  First  
 we review the geometric engineering of $N=2$ QFT$_4$
in type IIA superstring theory on Calabai-Yau threefolds. We then extend this 
to geometric
engineering of $N=1$ QFT$_4$ in M-theory on $G_2$ manifolds. These manifolds 
are 
given by K3 fibration  over    
three-dimensional spaces with  affine $ADE$ Dynkin geometries.
  Relating our construction to the related D6-brane scenario described
using the method of $(p,q)$ webs used in type II superstrings on Calabi-Yau 
threefolds, we  discuss the gauge group
and matter content of our $N=1$ QFT$_4$.

\section{  Generalities on   K3 surfaces with   $ADE$  singularities}
 Since  the geometric engineering method is based on the compactification on 
manifolds  with 
K3 fibration,  we  first briefly review some basic facts  about   $ADE$  
geometries of this 
fibration     
useful for  studying   supersymmetric gauge theories  embedded 
in superstrings and M-theory compactifications on 
 $G_2$ manifolds. 
 
 Roughly speaking, $ADE$ geometries are singular  four-dimensional 
 manifolds asymptotic to  $ R^4/ \Gamma$, where  $\Gamma $ is  a discrete 
subgroup 
of  $ SU(2)$.  The resolution of the singularities of these 
surfaces is nicely 
described  using the  Lie algebra structures.
The  well known singularities of these spaces are  as follows: 
\begin{enumerate}
  \item  Ordinary 
singularities classified 
by the finite  $ADE$  Lie algebras.\item Affine singularities classified by 
the affine $ADE$  Kac-Moody 
algebras. \end{enumerate}
  More general singularities  classified   by indefinte Lie 
algebra  have  been studied in \cite{ABS}.  The latter are at the basis  of 
the 
derivation of the 
new four-dimensional $N=2$  superconformal field theories.
  
  For simplicity, let us focus our attention on reviewing 
briefly the main lines of the finite   $ADE$  geometries.   Extended 
singularities can be found in \cite{ KMV,BFS, ABS}.\\
In $C^3$ parametrised by the local coordinates $x,y,z$, ordinary $ADE$ 
geometries 
are described 
by the following complex surfaces
\begin {equation}
\begin{array}{lcr}
 f(A_n): \quad xy+z^n=0\\
f(D_n): \quad x^2+y^2z+z^{n-1}=0\\
f(E_6):\quad  x^2+y^3+z^4=0\\
f(E_7):\quad  x^2+y^3+yz^3=0\\
f(E_8): \quad x^2+y^3+z^5=0.\\
\end{array}
\end{equation}
The $C^3$ origin, $ x=y=z=0$, is a singularity since $ f=df=0$. These 
geometries 
can be 'desingularized'  by deforming the complex structure of the 
surface or 
varying its Kahler structure. The two deformations are equivalent due to the 
self-mirror property of the  $ADE$  geometries . This is why we shall 
restrict 
ourselves  hereafter to giving only the Kahler deformation. The latter 
consists in 
blowing up the singularity by a collection of 
intersecting
 complex curves. This means that  we replace the singular point 
$(x,y,z)=(0,0,0)$ by a set 
of intersecting
 complex curves ${\bf P^1}$  (two-cycles). The 
nature of the set of
 intersecting ${\bf P^1}$ curves depends  on the type 
of the singular 
surface one is considering. The 
smoothed  $ADE$  surfaces
 share several features with the  $ADE$ Dynkin diagrams. In particular, the 
intersection matrix   of the complex 
curves used in the 
resolution of  the $ADE$  singularities is, up to some details, minus
the  $ADE$ Cartan matrix
 $K_{ij}$. This link leads to a nice correspondence between the  $ADE $ 
roots 
$ \alpha_i$  and  two-cycles involved in the deformation of the $ADE$ 
singularities. 
More 
specifically, to each 
simple root $ \alpha_i$, we associate a single  $({\bf P^1})_i$.
This nice connection between the geometry of  $ADE$  surfaces and Lie 
algebra turns out to 
be 
at the basis of important developments in  
string theory. In particular,  
the abovementioned link has been successfully used in:
\\(i) The geometric engineering of the  $ N=2$ supersymmetric  
four-dimensional quantum 
field theory 
embedded in type II superstrings on Calab-Yau  in the presence of D-branes 
\cite{KMV,BFS,ABS}.\\(ii) Geometric 
engineering of $ N=1$  quiver theories   using   conifold geometric 
transitions \cite{ V, 
AMV, CIV, CKV, CFIKV, AGMV}.
 
\section{ Geometric engineering of  $N=2 $ QFT in four dimensions  }
 In this section we review the geometric engineering method of $N=2 $ models  
embedded in type IIA superstring in four dimensions. Then we  extend this 
method  to  engineering $N=1$ models   in the context of  M-theory on  $G_2$  
manifolds.
\subsection{  $SU(2)$ Yang-Mills  in six 
dimensions} The main steps in getting  $N=2 $  four-dimensional QFT from type 
IIA  superstrings on Calabi-Yau threefolds  is  to start first  with the 
propagation of 
type IIA superstrings
on  K3 surfaces  in the presence of D2-branes wrapping on two-cycles. 
 Then we 
compactify the resulting model down to four dimensions. To illustrate the 
method, suppose  that K3  has an $su(2)$ singularity. In the vicinity of the 
$su(2)$ 
singularity, the fiber 
K3  may be described by the 
following equation
$$ xy=z^2,$$
where $x, y$ and $z$ are complex variables.  The deformation of this 
singularity consists in replacing the singular point 
$x=y=z=0$
 by a $\bf P^1$ curve parameterized by a new variable $ x'$ defined as 
$x'={x\over 
z}$.
 In the new local coordinates  $(x',y,z)$, the equation of the $A_1$ 
singularity may 
be written as:
\be
  x' y=z
\ee
which is not singular.
The next step is to consider the propagation of type IIA superstrings in this 
background  in the presence of a  D2-brane wrapping around the blown-up $\bf 
P^1$ 
 curve (real two-sphere)  parametrized by $x'$.  This gives  two 
$W_\mu^{\pm}$ 
vector particles, one for each of the two possible orientations for wrapping. 
These 
particles have 
mass proportional to the volume of the blown-up real two-sphere. 
$W_\mu^{\pm}$ are charged 
under the $U(1)$ field $Z_0^\mu$ obtained by decomposing the type IIA 
superstring three-form in 
terms of the harmonic form on the two-sphere. In the limit where the blown-up 
two-sphere $x'$ 
shrinks, we get three masslesss vector particles $W_\mu^{\pm}$  and 
$Z_0^\mu$ which form 
an $SU(2 )$ adjoint representation. We thus obtain an $N=2$ $SU(2)$ gauge 
symmetry in six
dimensions.
\subsection{ $N=2$  models in four dimensions}
 A further 
compactification on a  base $B_2$, that is on a real two-sphere, gives $N=2$ 
pure  $SU(2)$ Yang-Mills  in  four dimensions.  This  geometric $SU(2)$ gauge 
theory analysis can be easily extended to all simply-laced $ADE$ gauge groups. 
To 
incorporate matter, one 
should consider a non trivial geometry on the base $B_2$  of the  Calabi-Yau 
threefolds.  For 
example, if 
we have a two-dimensional locus with $SU( n)$ 
singularity and another locus with $SU( m)$ 
singularity and 
they meet at a point, the mixed wrapped two-cycles will now lead to $(n ,m) $ 
$N=2$ bi-fundamental
 matter of the $SU(n)\times SU(m)$ gauge symmetry in four dimensions. 
Geometricaly, this means 
that the base geomerty of  the  Calabi-Yau threefolds   is given by two 
intersecting $\bf  P^1$ curves, according to $A_2$ finite Dynking diagram,  
whose volumes $V_1$ 
and $V_2$  define the gauge coupling constants $g_1$ and $g_2$ of the $SU(n)$ 
and $SU(m)$ gauge
 symmetries, respectively. Fundamental matter is given by taking the limit 
$V_2$ to infinity, or 
equivalently $g_2=0$, so that the $SU(m)$  group becomes a flavor symmetry. 
Geometric engineering of the  $N=2$  four-dimensional QFT shows moreover that 
the 
analysis we have 
been describing
 recovers naturally some remarkable features which follows from the connection 
between  the resolution of singularities and Lie algebras. For instance, 
taking 
$m=n$ and identifying the 
$SU(m)$ gauge symmetry 
with $SU(n)$ by equating the $V_1$ and $V_2$ volumes, which imply in turn that 
$g_1=g_2$, the 
bi-fundamantal matter becomes then an adjoint one. This property is more 
transparent in the language 
of the representation theory.  The  adjoint of $SU(n+m)$ splits into 
$SU(n)\times SU(m)$ representations as:
\be
(n+m){\overline{(n+m)}}= n.{\bar n}+m.{\bar m} +{\bar n}.m+n.{\bar m},
\ee
 where $n.{\bar n}+m.{\bar m}$ gives the gauge fields and ${\bar n}.m+n.{\bar 
m} $ define the 
bi-fundamental matters.

We note that   $ N=2 $  four-dimensional QFT's 
are represented by quiver diagrams where to each 
$SU$ gauge group factor
 one associates a node and for each pair of groups with bi-fundamental matter, 
the two corresponding 
nodes are connected with a line. These diagrams are similar 
to the  $ADE$ Dynkin graphs. 
\section{$N=1$  $ADE$ quiver models from M-theory compactification}
 In this  talk  we discuss a straightforward way of elevating the geometric 
engineering of $N=2$ gauge models in type IIA superstring theory to a similar 
construction of $N=1$ $ADE$ quiver gauge models in M-theory 
\cite{BDR}\footnote{ 
Alternative studies of four-dimensional $N=1$ gauge models in the
framework of F- or M-theory may be found in  \cite{Va,ACHA}.}. 
The method we will be using here is quite similar to the   
geometric engineering of $N=2$ QFT$_4$.  Indeed, we 
 start with a local description of M-theory 
compactification on a $G_2$ manifold with K3 fibration over  
a real three-dimensional base space $B_3$. The 
scenario of type IIA superstring theory in six dimensions appears in seven 
dimensions in M-theory.  In this way, the  type IIA D2-branes are 
replaced by M2-branes, and we end with a   seven-dimensional pure  gauge  
models. 

To   get models with only  four supercharges in four 
dimensions, we  need to  compactify the  seven-dimensional 
model on a  three-dimensional base preserving  $1/4$ of the remaining 
16 supercharges. The internal space  must 
have vanishing first Betti number, $b_1=0$, to meet the requirement
of $G_2$ holonomy. An example of such a geometry has the 
three-dimensional sphere $\bf S^3$ as base.  However, in this case   we obtain 
only  {\em pure} $N=1$ Yang-Mills theory.   The incorporation of   
matter may be achieved by introducing a non-trivial geometry in the 
base of the K3 fibration. This leads us to consider
a three-dimensional intersecting geometry to describe a product
gauge group with bi-fundamental matter in four dimensions.  
\par 
  Our method  here is 
quite simple and motivated by the  work on Lagrangian sub-manifolds in 
Calabi-Yau 
manifolds \cite{GVW}.   To do so,  we consider the three-dimensional base space
as a two-dimensional fibration over a one-dimensional base, where 
the fiber and the base each preserves half of the 
seven-dimensional M-theory  supercharges. The entire base space $B_3$
could be  embedded in a three-dimensional complex Calabi-Yau  threefolds.
The latter  are realized explicitly as a family of (deformed)
$ADE$ singular K3 surfaces over the complex plane. In this way our   base 
geometry   can be identified with  Lagrangian sub-manifolds in such  
Calabi-Yau 
manifolds. It  is easy to  see that 
non-trivial three-cycles, satisfying the 
constraints of the $G_2$ base geometry, constitute  
$ADE$  intersecting  two-cycles of K3 surfaces fibered over a line segment  
in the complex plane.
\par  For simplicity,  let us consider  the case  of
$A_1$ singularity. This is subsequently extended
to more general $ADE$ geometries. The deformed $A_1$ geometry 
is given by 
\be
 z_1^2+z_2^2+z_3^2= \mu 
\label{mu}
\ee 
where $\mu$ is a complex parameter.  The $A_1$ 
threefolds may be obtained by varying the parameter $\mu$ over 
the complex plane parametrized by $w$
\be
 z_1^2+z_2^2+z_3^2= \mu(w).
\ee 
The base space $B_3$ can then be viewed as a finite line segment with an 
$\bf S^2$ fibration, where the radius $r$ of $\bf S^2$ vanishes at the two 
interval 
end points, and at the end points only \cite{HI}. The latter requirement
ensures that no unwanted singularities are introduced.
One way to realize this geometry is 
\be 
 r\ \sim\ \sin x
\label{r}
\ee
where $x$ is a real variable parameterizing the interval $\[0,\pi\]$ 
in the $w$-plane. 

This construction may be extended to more  complicated 
geometries where we have intersecting  spheres according to affine $ADE$
Dynkin diagrams.  This extension has a nice toric geometry 
realization,  where the  ${\bf S^2}$ can be viewed as a segment $[v_1,v_2]$ 
with 
a
circle on top, where it shrinks at the end points $v_1$ and $v_2$ \cite{LV}. On
the other hand, $\bf S^3$ can be viewed as an interval  with  two circles on 
top, where a  $\bf S^1 $  shrinks at the  first end  and the other $\bf 
S^1$ shrinks at the second end. In the resolved elliptic singularity,  $\bf 
S^1 \times S^1 $ can be identified with a collection of two-cycles $B_2$ 
according to affine  $ADE$ Dynkin diagrams. 
 In this way, the intersecting matrix of this geometry is given by the Cartan 
matrix of $ADE 
$ affine  Lie algebra.

\par Having  determined  the base geometry of the $G_2$ manifold with K3 
fibration, 
we  will discuss the corresponding gauge theory of the compactified
M-theory using a type IIA  superstring dual description. Indeed,  
M-theory 
on $G_2$ manifolds  can be related to  type IIA  superstrings  on Calabi-Yau 
threefolds in the presence of D6-branes wrapping Lagrangian sub-manifolds 
and filling the four-dimensional Minkowski space \cite{CSU}.  
For instance, a local description of M-theory near the $A_{n-1}$ singularity 
of K3 surfaces is equivalent to $n$ units of D6-branes \cite{LV}. 
Indeed, on the seven-dimensional world-volume of each D6-brane 
we have a $U(1)$ symmetry. When the $n$ D6-branes approach each other,
the gauge symmetry is enhanced from $U(1)^{\otimes n}$ to $U(n)$. 
An extra compactification of M-theory down to four-dimensional  space-time is 
equivalent to wrapping D6-branes on the same geometry.
In this way, the  D6-brane physics  can be 
 achieved  by  using the method of  $(p,q)$ webs  used in type II superstrings 
on 
toric Calabi-Yau manifolds
\cite{HI,FHHI,FH,U,CIM}. Using the results of this method,
we expect that the gauge model  in four-dimensional Minkowski space
has gauge group
\be
 G\ =\ \bigotimes_i U(n_i).
\ee 
The integers $n_i$ are specified by the anomaly cancellation 
condition. This means that they should form 
a null vector of the intersection matrix of the three-cycles $I_{ij}$.
In the infra-red limit the $U(1)$ factors decouple and one is left 
with the gauge group $ G\ =\ \bigotimes_{i=1}^m SU(n_i)$. The gauge group and 
matter content depend on the  intersecting geometry in the three-dimensional 
base of the $G_2$ manifold.  Identifying  the base with a collection of
the three-cycles    being  $\bf S^2$ fibrations over a line segment, the   
intersection matrix of 
the three-cycles  can be identified with the Cartan matrix, $K$, of 
the associated $ADE$ Lie 
algebra:
$
 I_{ij}=-K_{ij}
$. The anomaly cancellation condition is now translated into 
a condition on the affine Lie algebra, so the gauge group becomes  
$
 G\ =\ \bigotimes_{i} SU(s_in).
$ 
 where  $s_i$ are the corresponding Dynkin numbers. The resulting models are 
$N=1$ four-dimensional quiver  models with bi-fundamental matter. They are  
represented  by  $ ADE$ affine Dynkin diagrams. 
\\[.3cm] 
In this talk  we have discussed the geometric engineering of $N=1$
four-dimensional  quiver models. In particular, we have considered  models 
embedded in  M-theory on a $G_2$  manifold  with  K3 fibration over a 
three-dimensional base space with $ADE$ geometry.  This  base geometry is 
identified with 
 $ADE$ intersecting two-cycles over a line segment.  Using  the connection 
between M-theory and D6-brane physics 
of  type IIA superstring, we have given the physics content of 
M-theory compactified on such a $G_2$ manifold.  The corresponding  gauge 
model has been discussed in  terms 
of $(p,q)$  brane webs. 
\\[.3cm]
{\bf Acknowledgments.}
The author thanks   L.B. Drissi,
J. Rasmussen  and E.H. Saidi for collaborations
related to this work.   He is supported by  Ministerio de Educaci\'{o}n 
Cultura y 
Deportes  (Spain) grant  SB 2002-0036.

\end{document}